# Scenarios Generation- based Multiple Interval Prediction Method for Electricity Prices based on Scenarios Generation: Results

Xin Lu

**Abstract:** This paper introduces an innovative interval prediction methodology aimed at addressing the limitations of current evaluation indicators while enhancing prediction accuracy and reliability. To achieve this, new evaluation metrics are proposed, offering a comprehensive assessment of interval prediction methods across both all-sample and single-sample scenarios. Additionally, a novel Pattern-Diversity Conditional Time-Series Generative Adversarial Network (PDCTSGAN) is developed, designed to generate realistic scenarios and support a new interval prediction framework based on scenario generation. The PDCTSGAN model incorporates unique modifications to random noise inputs, enabling the creation of pattern-diverse and realistic scenarios. These scenarios are then utilized to produce multiple interval patterns characterized by high coverage probability and reduced average width. The proposed approach is validated through detailed case studies, and the paper concludes with a discussion of future research directions to further refine interval prediction techniques.

## 1. Introduction

The accurate forecasting of electricity prices has become a critical area of focus in recent years [1, 2], as it is essential for market participants to optimize profits, manage risks, and maintain stable operations [3, 4]. Electricity prices are subject to a wide range of influencing factors and are often regarded as anti-persistent data. These factors, particularly those related to electricity demand and renewable energy generation, lead to price fluctuations and pose significant challenges for accurate forecasting.

Electricity price prediction is generally categorized into point prediction and interval prediction. Point prediction employs deterministic approaches to estimate specific values by minimizing metrics such as root mean square error (RMSE) and mean absolute error (MAE) between the forecasted and actual data. However, the inherent uncertainties in electricity markets—arising from incomplete datasets or unexpected events—limit the reliability of point prediction methods. To address these limitations, interval prediction techniques have been developed. These methods provide a probabilistic framework to account for variability and enhance the robustness of predictions in the face of uncertainty.

The base model of this study is CTSGAN, which was published in [5]. After further improvements, this work has been published in [6, 7]. We applied this prediction model to the fields of batteries [8], virtual power plants [9, 10], and shared energy storage [11], achieving promising results in each application [12].

## 2. Case studies and results analysis

### 2.1 Implementation details

2.2 Evaluation of prediction intervals for each sample with PDCTSGAN

To evaluate the effectiveness of the prediction interval for each sample in the electricity price, Fig. 6 shows the prediction results of the day-ahead electricity price prediction intervals for 5 consecutive days. Fig. 6 contains 9 sub-figures, generated by CTSGAN after well-training, with the same conditions and different distribution random noise with 9 different $\sigma$ as input. Each sub-figure also contains 3 sub-subfigures, from top to bottom are the results of prediction intervals based on one prediction, the ECP and EAW of each point based on 100 predictions, respectively. As can be seen from the Fig. 6, the prediction results are relatively independent for each individual sample, with different ECPs and EAWs calculated by 100 prediction intervals. In general, the ECP is usually lower at the sharp fluctuations, and the EAW is larger at the spikes than at the troughs.

Fig. 6(c) gives the prediction results with standard Gaussian distribution as random noise input, from which it can be seen that there are some prediction intervals with low ECPs, mainly in two time periods, around time 85 with ECPs of approximately 10% and around time 160 with ECPs of almost 0. For the other samples, the ECPs are all above 80% and some are even 100%, meaning that 100 different prediction intervals all effectively contain each individual sample. If the random noise input of the CTSGAN is more concentrated, the occurrence of lower ECP becomes more frequent, with the occurrence of ECP=0 with $\sigma$=0.333 (Fig. 6(a)) being significantly higher than that with $\sigma$=0.666 (Fig. 6(b)) or $\sigma$=1 (Fig. 6(c)). In contrast, if the random noise input is distributed more widely, the CTSGAN will generate rich-diversity scenarios, achieving low-probabilities reinforcement prediction, and the interval will contain more possibilities. In Fig. 6(d) ($\sigma$=1.333), only the ECPs around time 160 are relatively low (about 0.2), and the ECPs around 85 are approximate larger than 95%, which are significantly larger than that in Fig. 6(c). With the further widening of the random noise distribution, each samples around time 160 can also be well predicted, and the ECP becomes progressively larger with $\sigma$. At $\sigma$=3, the ECPs round time 160 are larger than 80%, and for other samples, the ECPs are 100%.

One conclusion can be drawn that the prediction intervals for each sample are relatively independent, with some samples, for example, the samples around 160 in Fig. 6, being difficult to predict due to their own characteristics. The ECPs for all samples and for one sample are also independent of each other. Therefore, there is a problem with using the ECP for all samples to evaluate the ECP of one sample (LUBE method). Just as the overall ECP cannot be used to evaluate the validity of the prediction interval for one sample around 160.

Fig. 7 gives the ECP and EAW of 100 repeating prediction intervals for one sample at time 161, which is the worst prediction in Fig. 6. It is clear that ECP increases rapidly with $\sigma$ increasing and EAW increases slowly with $\sigma$ increasing. The introduction of a wider distribution of random noise allows for more diverse scenarios and better coverage of low probability sample in the prediction interval. Random noise input is fully trained in the CTSGAN training phase, so that even the introduction of large $\sigma$ noise does not significantly increase the EAW of the prediction intervals, with the EAW of 0.216 for $\sigma$=1 and the EAW of 0.237 for $\sigma$=3.

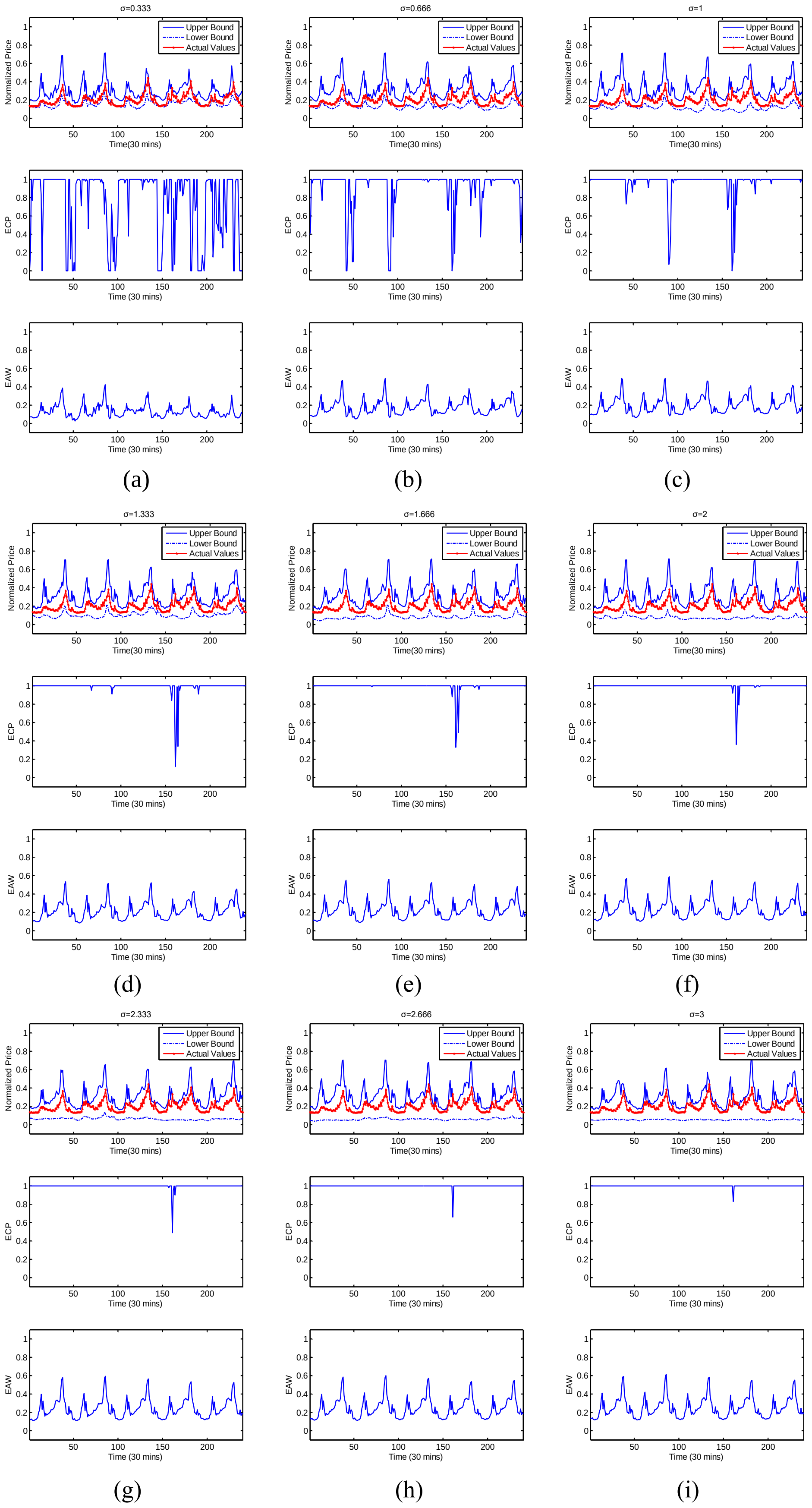

Fig.6 Prediction intervals of electricity price

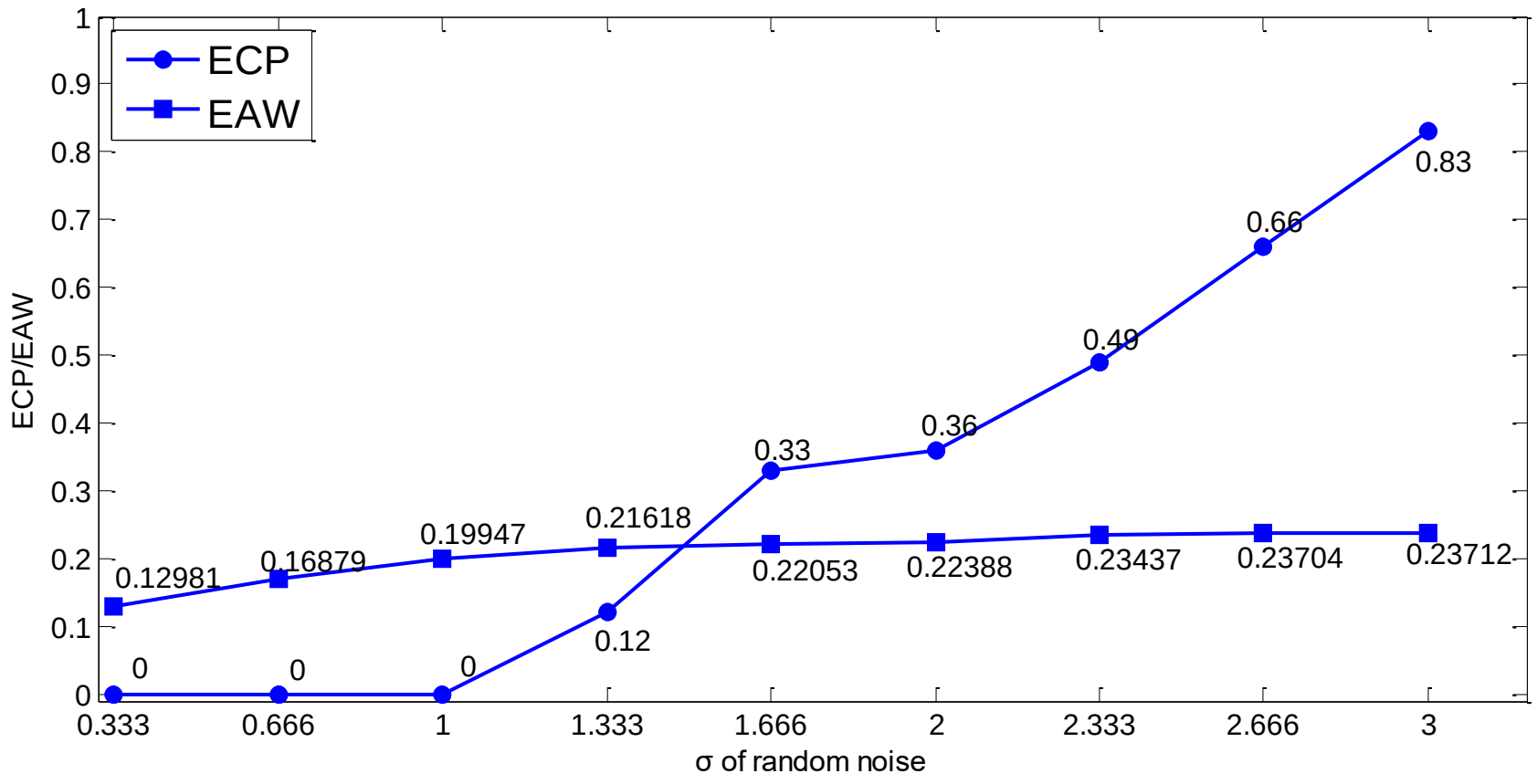

Fig. 7 ECP and EAW for the worst sample

### 2.3 Evaluation of prediction intervals for all samples with PDCTSGAN

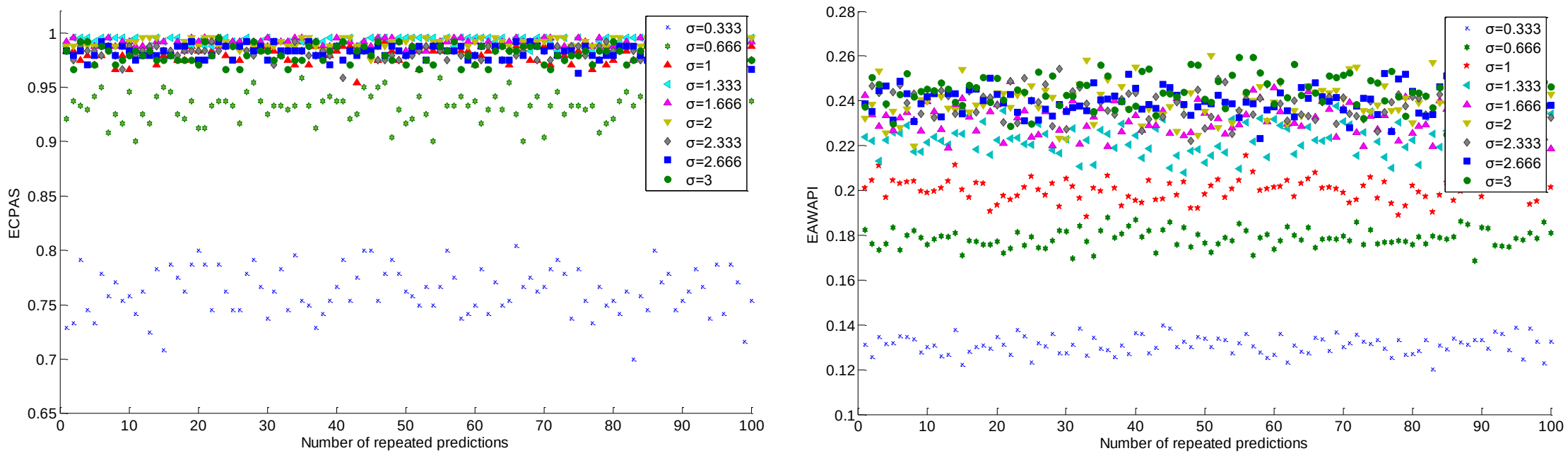

(a) (b)

Fig. 8 ECPAS and EAWAPI based on 100 repeated simulations

Day-ahead intervals prediction of electricity prices for all 5 consecutive days are conducted and repeated 100 times, and then the indicators of ECPAS and EAWAPI are calculated by (5) and (7) for each simulation. Fig. 8 gives the ECPAS and EAWAPI based on 100 repeated simulations. It can be seen that 100 repeated simulations have different prediction results due to different random noise input. A smaller $\sigma$ (smaller than 1) brings greater fluctuations in the indicator of ECPAS and poor ECPAS, and when $\sigma$ gradually increases, ECPAS gradually approaches 100%. EAWAPI also increases as $\sigma$ becomes larger, but the gain of ECPAS becomes smaller and smaller. This means that the proposed prediction intervals construction method based on scenarios generation will not expand the upper and lower boundaries indefinitely so that the prediction interval contains all samples. With ECPAS and EAWAPI for 100 predictions, the confidence level for ECPAS and EAWAPI can be obtained by (6) and (8).

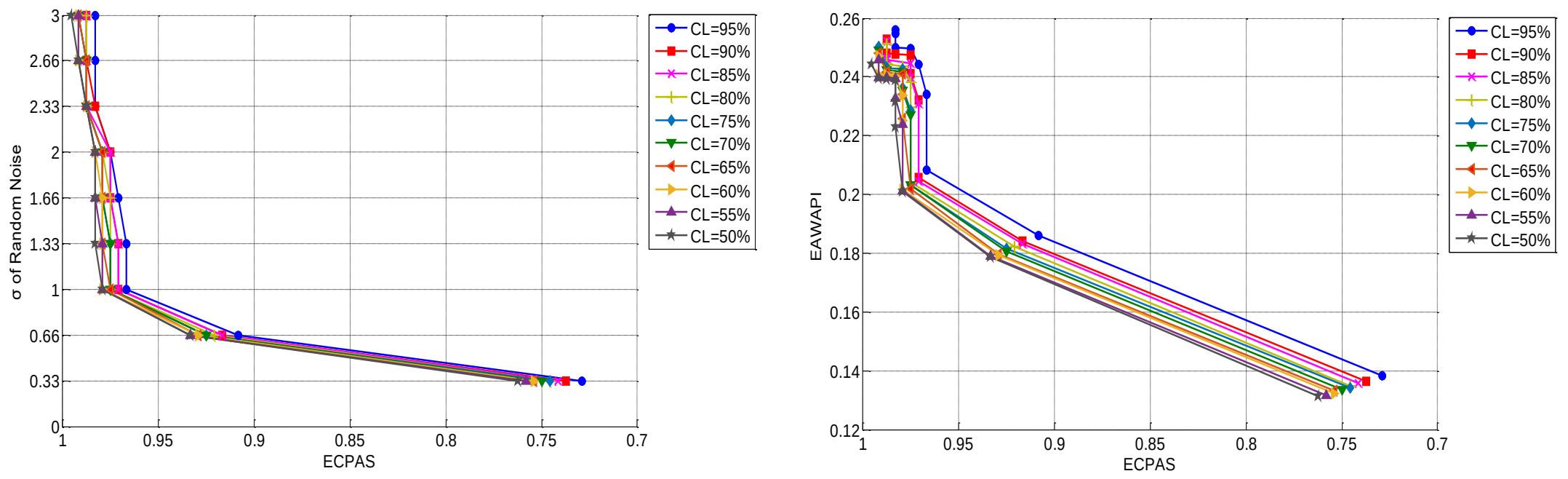

(a) (b)

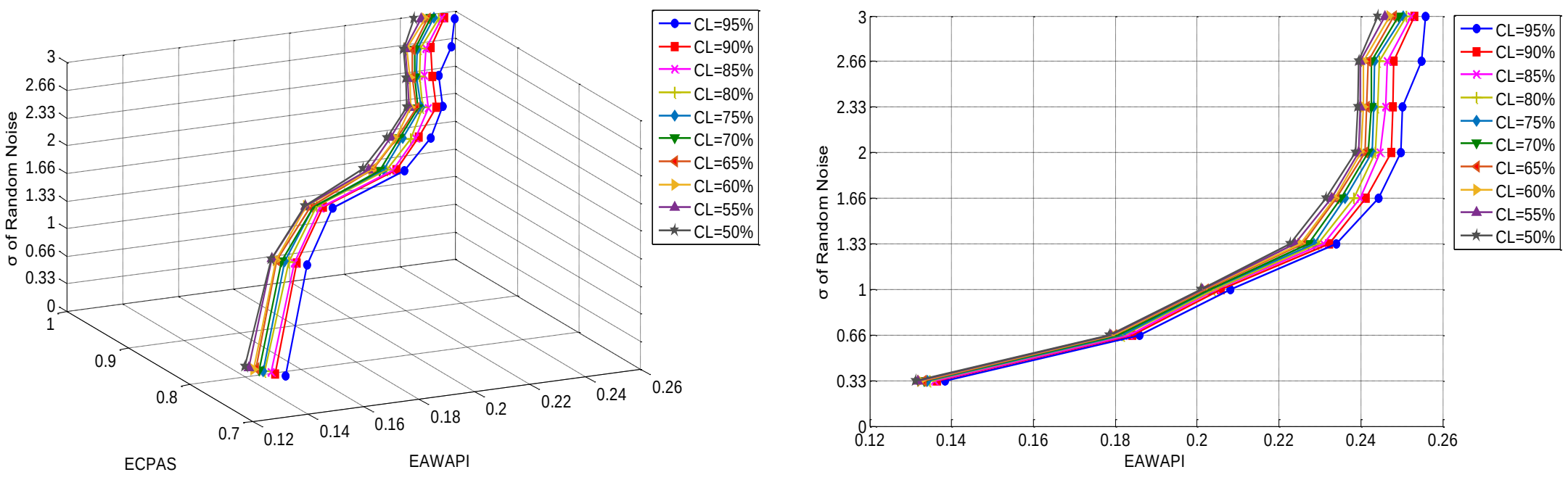

(c) (d)

Fig. 9 Confidence level for ECPAS and EAWAPI with different $\sigma$

Fig. 9(a) shows the values of ECPAS and EAWAPI with different confidence level and different $\sigma$ of random noise input, Fig. 9 (b-d) present the relationships between EAWAPI and $\sigma$, ECPAS and $\sigma$, and EAWAPI and ECPAS, respectively. As shown in Fig. 9(a), it is obvious that ECPAS and EAWAPI are different with different confidence levels, but ECPAS and EAWAPI change along with the change of $\sigma$ of random noise input, and the trends are almost the same. From Fig. 9(b), when $\sigma$=0.333, the difference of EAWAPI at different confidence levels is not very large, with 0.1384 (CL=95%) and 0.1312 (CL=50%). With the gradual increase of $\sigma$, the difference of EAWAPI at different confidence levels gradually increases, when $\sigma$ =3, EAWAPI=0.2559 (CL=95%) and 0.2441(CL=50%). If the confidence level is fixed, it can be seen that the increase in the input random noise distribution $\sigma$ leads to the increase in the prediction intervals width of all samples, for example, when confidence level is 90%, the EAWAPI is 0.1364 ($\sigma$=0.333) and 0.2530 ($\sigma$=3). In Fig. 9(c), contrary to the trend of EAWAPI, the variability under different confidence levels of ECPAS has not changed significantly with the increase of random noise input distribution $\sigma$. While $\sigma$ is 1, ECPAS is 0.9667 (CL=95%) and 0.9792 (CL=50%), and while $\sigma$ is 2, ECPAS is 0.9750 (CL=95%) and 0.9833 (CL=50%), and while $\sigma$ is 3, ECPAS is 0.9833 (CL=95%) and 0.9958 (CL=50%). When $\sigma$=2.333 or 2.666, ECPAS at different confidence levels overlap, for example, ECPAS remains at 0.9875, when CL is between 85% and 50% ($\sigma$=2.333), or between 90% and 65% ($\sigma$=2.666), which means that prediction intervals can cover 237 of 240 samples (such as Fig. 6(g) and (h)) and there are only three low-probability samples, and the current random noise distribution $\sigma$ cannot satisfy the prediction of these samples. In this case, if the concepts of ECP and ECPAS are confused, the conclusions brought are completely wrong. For example, each prediction interval has a probability of 98.75% to cover each sample. But in fact, except for these three low probability points, the coverage probability of each interval is 100%.

Fig. 9(d) shows the relationship between EAWAPI and ECPAS with confidence level from 50% to 95%. Overall, EAWAPI at any level of confidence increases with increasing ECPAS, and in addition, the incremental EAWAPI also increases as the ECPAS increases. The curves at different confidence levels are relatively evenly spaced, indicating that the prediction interval indicators calculated from proposed PSCTSGAN method are evenly distributed and can meet the confidence levels required for different usage situations.

## 3. Conclusion and further research

In this paper, firstly, a fallacy about the confidence level and coverage probability is pointed out, which is often used in the LUBE method to guide the construction of prediction intervals.

Secondly, to address the problem that current evaluation indicators for prediction intervals confuse the total samples with individual sample, this paper gives new evaluation indicators for one sample, ECP and EAW, and all samples, ECPAS and EAWAPI. Thirdly, a novel scenarios generation- based intervals prediction method is proposed. By changing the distribution of random noise input, the proposed PDCTSGAN method can give prediction intervals with different ECPAS and EAWAPI, meeting the diverse requirements in practical applications, such as in risk management of virtual power plant (VPP) or microgrid (MG).

Reference

[1] Fan S, Hyndman R J. The price elasticity of electricity demand in South Australia[J]. Energy Policy, 2011, 39(6): 3709-3719.

[2]Boland J, Filar J A, Mohammadian G, et al. Australian electricity market and price volatility[J]. Annals of Operations Research, 2016, 241: 357-372.

[3] Mwampashi M M, Nikitopoulos C S, Konstandatos O, et al. Wind generation and the dynamics of electricity prices in Australia[J]. Energy Economics, 2021, 103: 105547.

[4] Apergis N, Lau M C K. Structural breaks and electricity prices: Further evidence on the role of climate policy uncertainties in the Australian electricity market[J]. Energy Economics, 2015, 52: 176-182.

[5] Lu X, Qiu J, Lei G, Zhu J. Scenarios modelling for forecasting day-ahead electricity prices: Case studies in Australia. Applied Energy. 2022;308:118296.

[6] Lu X, Qiu J, Lei G, Zhu J. An interval prediction method for day-ahead electricity price in wholesale market considering weather factors. IEEE Transactions on Power Systems. 2023.

[7] Lu X, Qiu J, Yang Y, Zhang C, Lin J, An S. Large Language Model-based Bidding Behavior Agent and Market Sentiment Agent-Assisted Electricity Price Prediction. IEEE Transactions on Energy Markets, Policy and Regulation. 2024.

[8] Lu X, Qiu J, Lei G, Zhu J. State of health estimation of lithium iron phosphate batteries based on degradation knowledge transfer learning. IEEE Transactions on Transportation Electrification. 2023;9:4692-703.

[9] Lu X, Qiu J, Zhang C, Lei G, Zhu J. Assembly and competition for virtual power plants with multiple ESPs through a “recruitment–participation” approach. IEEE Transactions on Power Systems. 2023.

[10] Lu X, Qiu J, Zhang C, Lei G, Zhu J. Seizing unconventional arbitrage opportunities in virtual power plants: A profitable and flexible recruitment approach. Applied Energy. 2024;358:122628.

[11] Lu X, Qiu J, Zhang C, et al. Promoting Shared Energy Storage Aggregation among High Price-Tolerance Prosumer: An Incentive Deposit and Withdrawal Service[J]. arXiv preprint arXiv:2501.04964, 2025.

[12] Yan G, Trück S. A dynamic network analysis of spot electricity prices in the Australian national electricity market[J]. Energy Economics, 2020, 92: 104972.

[13] Can W, Ming N, Yonghua S, Zhao X. Pareto Optimal Prediction Intervals of Electricity Price. IEEE transactions on power systems. 2017;32:817-9.

[14] Zhou W, Lu X, Zhao D, Jiang M, Fan L, Zhang W, et al. A dual-labeled dataset and fusion model for automatic teeth segmentation, numbering, and state assessment on panoramic radiographs. BMC Oral Health. 2024;24:1201.

[15] Chen J, Shen Y, Lu X, Ji Z. A multi-Objective intelligent optimization prediction method for

wind power probability interval [J]. Power Grid Technol. 2016;40:2281-7.
[16] Chen J, Shen Y, Lu X, JI Z. An intelligent multi-objective optimized method for wind power prediction intervals. Power system technology. 2016;40:2758-65.
[17] Lu X, Yan X, Chen J. Ultra-short-term wind power intervals prediction considering randomness of wind power generation. ActaEnergiae Solaris Sinica. 2017;38:1307-15.
[18] Neyman J. Outline of a Theory of Statistical Estimation Based on the Classical Theory of Probability. Philosophical transactions of the Royal Society of London Series A: Mathematical and physical sciences. 1937;236:333-80.
[19] Morey RD, Hoekstra R, Rouder JN, Lee MD, Wagenmakers E-J. The fallacy of placing confidence in confidence intervals. Psychonomic bulletin & review. 2016;23:103-23.
[20] Forrest S, MacGill I. Assessing the impact of wind generation on wholesale prices and generator dispatch in the Australian National Electricity Market. Energy policy. 2013;59:120-32.
[21] Yoon J. End-to-End Machine Learning Frameworks for Medicine: Data Imputation, Model Interpretation and Synthetic Data Generation. ProQuest Dissertations Publishing; 2020.
[22] Arjovsky M, Chintala S, Bottou L. Wasserstein generative adversarial networks. International conference on machine learning: PMLR; 2017. p. 214-23.
[23] Durugkar I, Gemp I, Mahadevan S. Generative multi-adversarial networks. arXiv preprint arXiv:161101673. 2016.
[24] Ghosh A, Kulharia V, Namboodiri VP, Torr PH, Dokania PK. Multi-agent diverse generative adversarial networks. Proceedings of the IEEE conference on computer vision and pattern recognition2018. p. 8513-21.
[25] Che T, Li Y, Jacob AP, Bengio Y, Li W. Mode regularized generative adversarial networks. arXiv preprint arXiv:161202136. 2016.
[26] Padala M, Das D, Gujar S. Effect of Input Noise Dimension in GANs. arXiv preprint arXiv:200406882. 2020.
[27] Mao Q, Lee H-Y, Tseng H-Y, Ma S, Yang M-H. Mode seeking generative adversarial networks for diverse image synthesis. Proceedings of the IEEE/CVF Conference on Computer Vision and Pattern Recognition2019. p. 1429-37.
[28] Shen Y, Lu X, Yu X, Zhao Z, Wu D. Short-term wind power intervals prediction based on generalized morphological filter and artificial bee colony neural network. 2016 35th Chinese control conference (CCC): IEEE; 2016. p. 8501-6.
[29] Kavousi-Fard A, Khosravi A, Nahavandi S. A New Fuzzy-Based Combined Prediction Interval for Wind Power Forecasting. IEEE transactions on power systems. 2016;31:18-26.